\documentclass[epj,nopacs,final]{svjour}
\usepackage{graphics}
\usepackage{amsfonts}

\def\s2{{\mathcal S}_2} 
\def\wr{{\mathcal W}r}
\def\ts{{\bf t}(s)}
\def\n{{\bf n}} 
\def\r{{\bf r}} 
\def\t{{\bf t}} 
\def\b{{\bf b}} 
\def\K{ {\cal K} }
\def\ez{{ {\bf \hat e}_z} }

\begin{document}
\title{Writhing Geometry of Stiff Polymers and Scattered Light }
\author {A.C. Maggs} \institute{Laboratoire de Physico-Chimie
  Th\'eorique, ESPCI-CNRS, 10 rue Vauquelin, 75005 Paris, France.
}
\date{12:52 14/9/2001} \abstract{The geometry of a smooth line is
  characterized {\sl locally}\/ by its curvature and torsion, or {\sl
    globally} by its writhe. In many situations of physical interest
  the line is, however,  not smooth so that the classical
  Frenet description of the geometry breaks down everywhere.  One
  example is a thermalized stiff polymer such as DNA, where the shape
  of the molecule is the integral of a Brownian process. In such
  systems a natural frame is  defined by parallel transport.
  In order to calculate the writhe of such non-smooth lines we study
  the area distributions of random walks on a sphere.  A
  novel transposition of these results occurs in multiple light
  scattering where the writhe of the light path gives rise to a Berry
  phase recently observed in scattering experiments in colloidal
  suspensions.}
\PACS{
      {}{}  
     } 
\maketitle
\section{Introduction}
The classic mathematical description of the geometry of a line in
three dimensional space uses a specific choice of frame, the Frenet
frame \cite{frenet} involving the tangent, $\t$ the normal, $\n$ and
binormal, $\b$ to the curve. However \cite{bishop} this description is
only valid for curves which are ${\mathcal C}^3$, and for which the
curvature, $\kappa$ does not vanish. Locally, by appropriate choice of
axes, the shape of a three dimensional curve is approximated by the
curve
\cite{localbook} as
\begin{equation}
{\bf r}(s) = \left ( s, {\kappa \over 2}s^2 , {\kappa \tau  \over 6} s^3 \right )
\end{equation}
where $\tau$ is the torsion and $s$ the curvilinear distance, showing,
explicitly, the need for high order derivatives in order to define the
torsion.  In this article we show that while many problems involving
elastic beams often fall within this class of smoothness the case of
{\sl non-smooth}\/ curves is very far from being a mathematical curiosity; it
is even the generic case when one is is interested the {\sl
  statistical mechanics}\/ of fluctuating lines. Recent
micromanipulation experiments performed on DNA molecules are one place
where these considerations have been found to be important.

Very similar ideas have recently been shown to apply in multiple light
scattering \cite{vincent} where the tortuosity of the light path
through a sample gives rise to a Berry phase. We shall show here that
the theory of Berry phases in multiple scattering samples is closely
related to the statistical mechanics of DNA considered as a
fluctuating, stiff thread.

\section{Intrinsic Geometry a Line}
The description of the geometry based on the Frenet frame can be
replaced by one based on the intrinsic geometry of the tangent space
to a curve: The unit tangent $\ts$ to a curve lives on a sphere,
$\s2$.  We now chose an arbitrary initial vector $\n' (0)$,
perpendicular to $\t(0)$. Clearly any vector which is perpendicular to
$\ts$ is in the tangent space of $\s2$, allowing a definion $\n'(s)$
by parallel transport from $\n'(0)$ \cite{acm}.  Finally, define $\b'$
to be perpendicular to both $\t$ and $\n'$, forming the alternative
frame. Clearly this recipe is defined for all curves for which $\ts$
is continuous even if non-smooth.

Parallel transport about a closed loop on a sphere leads to an
anholonomy.  This anholonomy has been independently discovered many
times in different fields of physics. Indeed Foucault's pendulum can
be understood as one simple example of this phenomenom \cite{book}.
In the literature on DNA and knots following Fuller \cite{fuller} and
White \cite{white} the anholonomy is known as {\sl writhing}, $\wr$;
it was discovered as part of a topological invariant of closed
ribbons, the linking number.  In the quantum literature the anholonomy
is the origin of the geometric phase discussed by Rytov \cite{rytov}
and Berry \cite{berry}.  Some physical consequences are illustrated in
a simple example in Figure (1). In this figure a square prism is 
folded into a non-planar configuration. One face of the prism is
marked with a vector ${\bf E}$. We see that the bending of the prism
in three dimensional space has led to a rotation of the vector ${\bf
  E}$ about the vertical axis.  We now consider the spherical curve
$\ts$ corresponding to Figure (1); application of the Gauss-Bonnet
theorem to $\ts$ allows one to calculate the angle of rotation: It is identical
to the area enclosed by $\ts$ on $\s2$.  Indeed we find that 
$\ts$ encloses $1/8$ of a sphere, corresponding to a rotation of
$\pi/4$ of the vector ${\bf E}$.

We note that for smooth lines for which $\ts$ is a closed curve the
local measure of the tortuosity of a curve, $\tau$ and the global
measure, $\wr$ are linked by the equation
\begin{equation}
2 \pi \,  \wr + \int \tau\, ds=0 \quad {\rm mod} \, 2 \pi.
\end{equation}.

\section{Statistical mechanics of bending of stiff polymers}
The bending energy of a stiff beam  in the slender body
approximation of elasticity is given by
\begin{equation}
E = {\K\over 2} \int \left( {d \t (s)\over ds}  \right)^2 \ ds
\label{stiff} \ ,
\end{equation}
where $\K$ is the bending modulus. Such elastic descriptions are often
used to describe the mechanics of stiff biopolymers.  At a non-zero
temperature the polymer retains memory of its orientation over a
characteristic distance $\ell_p=\K/k_B T$ where $T$ is the
temperature. $\ell_p$ is known as the {\sl persistence length}. This
length should be compared with the diameter, $d$ of the polymer.
Clearly it is only when $l_p/d$ is large that a continuum description
in terms of an elastic line has sense. For DNA one finds $\l_p \sim 50
nm$ and $l_p/d \sim 30$. Other stiffer biopolymers such as actin
filaments are also studied in the laboratory with values of $l_p/d$
 as large as $2000$. These filamentary systems are particularly
easy to study using methods such as scanning fluorescence microscopy
which allow direct measures of the three dimensional shape of the
system.

To calculate the partition function one must now sum over all paths
\begin{equation}
{\mathcal Z} = \sum_{paths} e^{-E/k_B T}
\label{pathint}
\end{equation}
The sum for the partition function is clearly closely related to path
integrals studied in quantum mechanics.  Formally the energy Eq.
(\ref{stiff}) looks like the kinetic energy of a free particle moving
on a sphere. From the path integral Eq. (\ref{pathint}) one derives a
Fokker-Planck equation which is entirely analogous to the Schroedinger
equation for a particle on a sphere:
\begin{equation}
{\partial P(\t ,s) \over \partial s}
=
{1\over 2 \ell_p}
\nabla^2P(\t ,s) \ ,
\label{fp}
\end{equation}
where $\nabla^2$ is the Laplacian operator on the sphere.  Here
$P(\t,s)$ is the probability of finding the chain oriented in the
direction $\t$ at the point $s$.  As a function of $s$ the vector
${\bf t}$ ``diffuses'' with diffusion coefficient $1/(2\ell_p)$. The
full distribution function of the orientational and spatial degrees of
freedom for the polymer $Q(\r,\t,s)$ is then found by noting that 
the shape of the polymer is found by integration of $\t(s)$ so that
\begin{equation}
{\partial Q \over \partial s}  + {\t}. \nabla_{\r} Q
=
{1\over 2 \ell_p}
\nabla_{\bf t}^2Q
\end{equation}
{\it i.e.}\/ the shape evolves by convection along $\t$.

From the diffusion like Eq. (\ref{fp}) we understand why the torsion
and the Frenet frame are not useful at non-zero temperatures: A
typical configuration of a stiff polymer is the realization of an {\sl
  orientational diffusion process}. $\ts$ is continuous but not
differentiable; the Frenet frame is undefined everywhere (with
probability 1). Despite this low smoothness it is possible to
study the parallel transport of the vector $\n'$.  

In applications in polymer physics one is interested in the magnitude
of torsional fluctuations (such as shown in Figure (1)) due to the
thermal fluctuations.  Using the Fuller-Berry result linking
anholonomy to $\ts$ one is lead to study the distribution distribution
of the area enclosed by loops by Brownian paths. This is in fact an
old problem, first treated by Levy \cite{levy} in the Euclidean plane
and more recently on the sphere \cite{orsay}.  For short chains we use
Levy's results to explicity find the the distribution of the angle of
rotation $\Phi$ due to writhing between the ends of a filament of
length $L$ as
 \begin{equation}
{\cal P} (\Phi)={\ell_p \over  2L } {1 \over \cosh^2(\Phi \,  \ell_p/L )}
\quad .
\label{writhedis}
\end{equation}
For long chains the problem is difficult, there are ambiguities in the
definition of area on the sphere, \cite{comment}. The Gauss-Bonnet
theorem only allows a definition of the area ${\rm mod}\, 4\pi$ and
the more elaborate expression of C\u{a}lug\u{a}rean based on the Gauss
linking linking number must be used to calculate the rotation angle.

\begin{figure}
  \resizebox{0.4\textwidth}{!}{\includegraphics{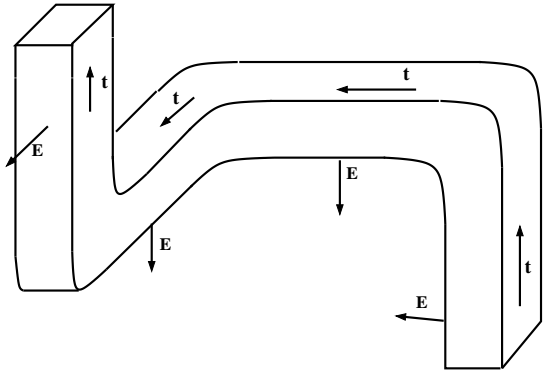} }
\caption{A tortuous bar in space can represent either the shape
  of a fluctuating molecule such as DNA, or the path of a light beam
  transmitted along a fibre. The relative rotation,
  $\Phi = \pi/2$, between the two ends is an example of {\sl writhe},
  $\wr$ where $\Phi=2 \pi\, \wr$.  For light transmitted along a fibre
  with polarization vector ${\bf E}$ this rotation is a simple example
  of a Berry phase.}
\label{fig:1}      
\end{figure}

\section{Application to DNA}
Very elegant experiments have been performed \cite{bensimon}
on the mechanics of thermalized DNA molecules and interpreted with
theories of fluctuating elastic threads. In these experiments a bead
is attached to each end of a long molecule. The beads are held in
magnetic traps which allows the simultaneous application of a force
and couple. To describe these experiments theoretically one adds the
force and couple into the bending energy Eq. (\ref{stiff}).  This
leads to a Hamiltonian analogous to that of a particle in the presence
of a Dirac monopole \cite{mezard}.
\begin{equation}
E = {\K\over 2} \int \left( {d \t (s)\over ds}  \right)^2  +
f \,  \t. \ez +
 \Gamma \, {(\ez \times \t)\over 1+ \t .\ez}. \dot \t \quad ds,
\label{dirac}
\end{equation} where $f$ is the external force pulling in the direction $\ez$
and $\Gamma$ is the external torque. Experimentally one measures the
mean separation of the two beads as a function of $f$ and $\Gamma$.
Good agreement is found with calculations based on Eq. (\ref{dirac}).
Most experiments can however be explained with a version of this
Hamiltonian expanded to quadratic order in the fluctuations in $\t$,
valid when the external force is larger than $k_BT/\ell_p$
\cite{nelson}.

Under large forces this continuum description of DNA molecules breaks
down due to an instability in the base structure of the molecule. For
even higher forces the hydrogen bonds holding the two DNA strands are
torn apart and the molecule is denatured. These events are clearly
beyond any continuum theory and are must be treated by atomistic
simulations \cite{richard}.

\section{Multiple light scattering}

If one shines light into a inhomogeneous medium then the direction of
propagation of the light is modified by scattering. Eventually the
light gets back to the surface and re-escapes from the material. Such
strongly scattering media are familiar from every day life, for
instance milk, white paper or even biological tissues. One is often
interested in optical imaging as deeply as possible in these materials: A
particularly important application is the diagnostics of severe burns;
one needs to know the degree of tissue damage as a function of depth
over a large area.  Empirically it is known that the quality of
imaging in such media is enhanced if polarization discrimination is
used to filter the light. Rather surprisingly circularly and linearly
polarized light do not display the same quality and resolution in
imaging.  We shall now show that the statistical mechanical treatment
of scattering in such media  is very close to that of the stiff
polymer introduced above.  The idea of parallel transport will be used
to understand the polarization patterns observed at the surface of
samples.

In systems with weak, large scale heterogeneities, the scattering of a
beam of light is largely in the forward direction, \cite{mackintosh}.
Each independent scattering event changes the direction of propagation
by some small random angle.  We model this process as giving rise to
angular diffusion in the propagation direction of the light just as in
Eq. (\ref{fp}).  As shown in \cite{goro} the helicity of photons is
conserved in forward scattering processes over a length which is
larger than $\ell_p$. We shall thus neglect such events in the
following discussion. This conservation implies that the polarization
vector of linearly polarized light evolves by parallel transport on
$\s2$ with $\t$ the direction of propagation of the light and $n'$ the
polarization vector during the scattering process.  This is exactly
the condition needed for the development of a Berry or geometric phase
in the scattered light.

\begin{figure}
  \resizebox{0.4\textwidth}{!}{\includegraphics{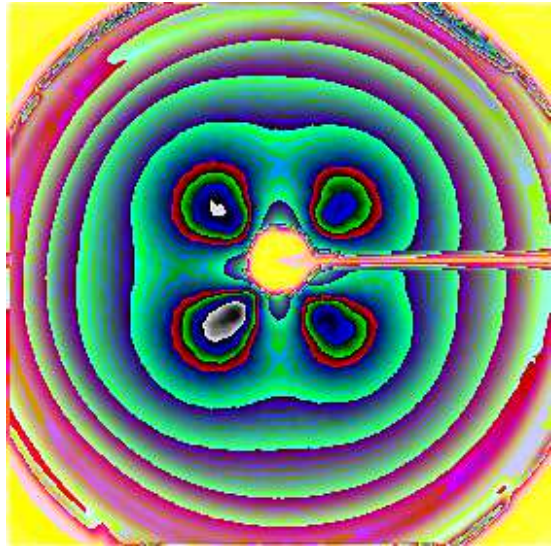}}
\caption{False color image of polarization patterns on the surface of a
  beaker of a colloidal solution {\it \cite{mueller}}\/ of $2 \mu m$
  latex beads. Linearly polarized light is incident at a spot at the
  center of the image.  The surface of the beaker is imaged with a
  camera through a linear analyzer. The analyzer is crossed with
  respect to the polarization of the incident beam.  We see a clear
  fourfold symmetry in the intensity of the backscattered light with
  the maximum in inclined intensity at an angle of $45$ degrees to the
  direction of the polarizers. }
\label{image}
\end{figure}

We use these remarks to understand recent rather elegant
experiments \cite{mueller} performed on colloidal solutions, motivated
by problems in biological imaging.  In the experiments a linearly
polarized light beam is focused to a point on the surface of a
colloidal solution. The surface of the solution is then imaged through
a second linear analyzer. One observes a clear pattern of brightness
on the surface with fourfold symmetry about the incident beam, Figure
(\ref{image})

\begin{figure}[ht]
  \resizebox{0.4\textwidth}{!}{\includegraphics{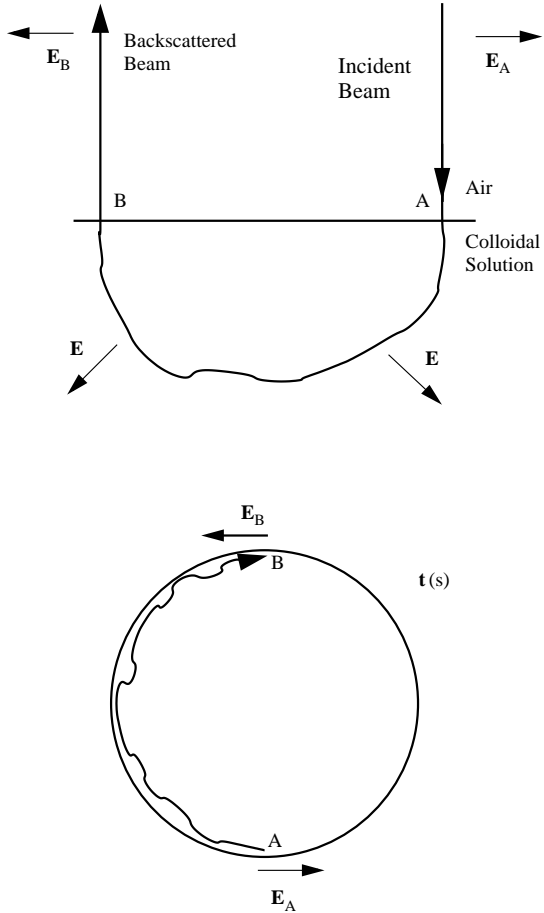}}
\caption{
  {{\sl top:}\/ Light is incident on a multiple scattering medium. The
    direction of propagation is randomly bent back and the light
    escapes from the surface.  {\sl bottom} } The direction of
  propagation of the scattered light is plotted on a sphere. Incident
  light corresponds to the south pole, $A$. The escaping light
  corresponds to the north pole, $B$. The indicated path is scattered
  principally to the left so that on the sphere the path remains, on
  average on the western area of the sphere. As the point of
  observation goes around the incident beam, an angle of $2\pi$, the
  path between the poles sweeps out an area on the sphere of $4\pi$.
  From the results on geometic phases we deduce that the polarization
  also rotates $4\pi$. }\label{backscatter}
\end{figure}

Let us proceed by translating the backscattering geometry into an
ensemble of paths on the unit sphere, $\{\t \}$ in order to apply the
Fuller-Berry's result.  As shown in Figure (\ref{backscatter})
backscattered light corresponds to a path from the south to north
poles of the sphere, $\ts$, describing the direction of propagation.
Each path from the south to north pole is a realization of a random
walk on the sphere.  We take as a reference state light scattered to
the left, polarized in the plane of the page.  For the reference state
the original polarization vector ${\bf E}_A$ is parallel transported
around the sphere, Figure (\ref{backscatter}, bottom) so that the
initial and final vectors are antiparallel.  Consider, now, a second
azimuthal direction with a trajectory ${\bf t'}(s)$.  This new
trajectory together with the reference path form a closed loop on the
sphere which allow us to apply the Berry result.  As the point of
observation on the sample, $B$, changes and winds an angle of $2\pi$
about the incident beam, $A$, the new path $ {\bf t'}(s) $ on the
sphere sweeps out a solid angle of $4 \pi$.

Clearly in a given azimuthal direction different paths are possible
with a distribution of values possible for the writhe. With an
incoherent light source one must sum the intensity over all paths in a
given azimuthal direction. Since these paths are correlated we deduce
that the polarization at the surface of the sample rotates $4 \pi$ or
{\sl two full turns}\/ as we move just once about the incident beam.
Since a linear analyzer is sensitive to the angle of rotation modulo
$\pi$ we see that there are four radial directions in which an
analyzer detects a maximum in the intensity. 

One can now understand that the difference in the coupling of the
Berry phase to linearly and circularly polarized light is partially
responsible for their different imaging qualities in tissue: An object
hidden deep under the surface of the beaker in Figure (\ref{image})
can only weakly modify the intensity at the surface. This weak
modification is easily hidden by the strong variation in intensity due
to the Berry phase.

\section{Conclusions}
We have seen that two rather different physical systems are expained
using simple ideas from the geometry of lines. In many physical
situations a description in terms of parallel transport and intrinsic
geometry is more natural than the classic description based on the
Frenet Frame.

Finally we note there is subtle point that we have skipped over in the
light scattering problem: The scattering cross-section of
spherical particles in the Rayleigh-Gans regime decays rather slowly
at large angles. The mean square scattering angle thus contains a
logarithmic divergence which implies that the angular diffusion
coefficient is not defined unless a cut off is introduced in the
problem. In systems with smoother variations of the optical properties
such as a critical fluid the scattering cross section falls off
quickly and no cut off is needed. This divergence is also linked with
the helicity filiping which occurs in scattering from spherical
particles \cite{goro}.

\vskip 0.5cm { The image of figure (2) was kindly provided by
  A.~H. Hielscher, Depts. of Biomedical Engineering and Radiology,
  Columbia University, New York, NY.

\end{document}